# Anomalous Features in Surface Impedance of Y-Ba-Cu-O Thin Films: Dependence on Frequency, RF and DC Fields

Anton V. Velichko, Adrian Porch, and Richard G. Humphreys

***Abstract*—Two high-quality Y-Ba-Cu-O thin films on MgO substrates have been investigated using the coplanar resonator technique at 8 and 16 GHz. Both films exhibit an anomalous decrease in their surface impedance, $Z_s$ as a function of microwave field, $H_{rf}$. In zero dc field, $H_{rf}$-dependences of $R_s$ and $X_s$ for both the samples are uncorrelated, and only one of the quantities, $R_s$ or $X_s$, displays anomalous behavior. Here, application of relatively weak (~5 mT) dc magnetic fields, $H_{dc}$ can produce a correlated decrease of $R_s(H_{rf})$ and $X_s(H_{rf})$. The dependences of $Z_s$ on $H_{dc}$ in both low and high microwave power regimes were found to be non-monotonic. The frequency dependence of $R_s \sim f^n$, $1.7<n<2.5$, remained the same upon the transition from low to high microwave power ranges. The consequences of the reported findings for microwave device applications are briefly discussed.**

## I. Introduction

LATELY several research groups have reported an anomalous decrease in the surface impedance, $Z_s = R_s + j \cdot X_s$, of high-quality HTS thin films as a function of dc, $H_{dc}$ [1]-[3] and microwave, $H_{rf}$ [2], [4], [5] magnetic fields. Though similar effects have been observed in low-$T_c$ superconductors (see [6], [7], e.g.), the peculiar nature of superconductivity in the cuprates (*d-wave* or mixed pairing mechanism), as well as the extremely short coherence lengths, anisotropy and complicated charge transfer may be responsible for different mechanism(s) of the observed anomalies. The field-dependent surface barrier mechanism [6] is not believed to be of particular importance to the observed anomalous behavior in HTS films [1]-[5], since the majority of the aforementioned experiments were performed with both dc and microwave magnetic fields perpendicular to the film surface. Therefore, magnetic flux entry in this case should not be impeded by the surface barrier. Another classical mechanism, such as stimulation of superconductivity by microwave radiation [8], is also irrelevant to the observations reported here, since according to Eliashberg [8], a dc magnetic field always suppresses the stimulation effect, whereas in our experiments the static field may enhance the anomaly in $Z_s(H_{rf})$.

Recently, two alternative mechanisms [9]-[11] have been suggested to explain the unusual behavior of $Z_s(H_{rf})$. One is based on the idea of the existence of grain-shunted weak links in high-quality HTS thin films [9], [10]; the other appeals to the possibility of the recovery of superconductivity in the cuprates by application of external (dc or rf) magnetic fields as a result of the alignment of the spins of uncompensated magnetic moments, which are believed to be present in non-optimally doped HTS materials [11].

In this paper we present some further results on $Z_s(H_{rf})$ for two high-quality epitaxial Y-Ba-Cu-O thin films studied at different frequencies (8 and 16 GHz) and in various (up to 12 mT) dc magnetic fields using the coplanar resonator technique [12].

## II. Experimental Results and Discussion

### A. Samples

The two film studied here, TF1 and TF2, are deposited by the e-beam co-evaporation technique onto polished (001)-oriented MgO single crystal substrates. The films are 350 nm thick. Typical c-axis misalignment was less than 1 %. Critical temperatures and critical current densities (at 77 K) are 83 K and 0.5 MA/cm$^2$, and 88 K and 2.2 MA/cm$^2$ for TF1 and TF2, respectively. The low microwave power values of the surface resistance, $R_s$ and penetration depth, $\lambda$ are 35 μΩ and 210 nm, and 50 μΩ and 135 nm, for samples TF1 and TF2, correspondingly. The absolute values of $\lambda$ were determined without resorting to any theoretical model by using the technique developed by Porch et al. [12]. The method is based on measuring the microwave response of two coplanar resonators patterned on the same film. The two resonators have different ground plane-to-central strip spacing, but the same resonant frequency. This allows us to extract absolute value of l at a fixed temperature, $T_{min}$ by adjusting the guess value for l($T_{min}$) in order to make the temperature dependences of $\lambda$ for the two resonators coincide over the whole *T*-range. This method gives one a

Manuscript received August 18, 2000. This work was supported in part by the EPSRC rolling Grant No. RRHA04844

A. V. Velichko is with the School of Electronic and Electrical Engineering, University of Birmingham, Birmingham B15 2TT, UK, on leave from the Institute for Radiophysics and Electronics of NAS, Kharkov 310085, Ukraine (e-mail: velichko@eee-fs7.bham.ac.uk).

A. Porch is with the School of Electronic and Electrical Engineering, University of Birmingham, Birmingham B15 2TT, UK (e-mail: a.porch@bham.ac.uk).

R. G. Humphreys is with DERA, Malvern, Worscester WR14 3PS, UK (e-mail: rghumphreys@dera.gov.uk).

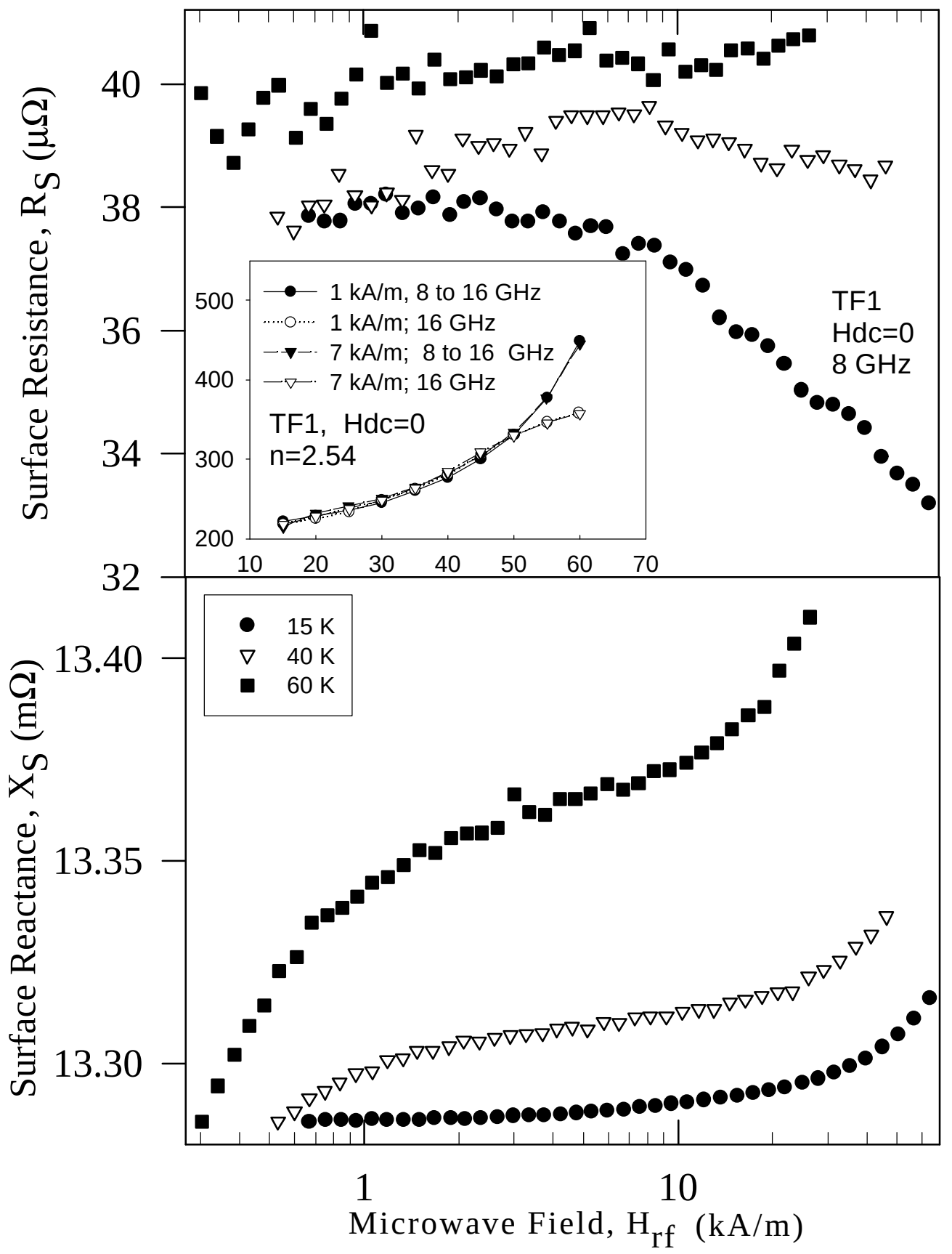

Fig. 1. Surface resistance, $R_s$ (a) and surface reactance, $X_s$ (b) as a function of microwave magnetic field, $H_{rf}$ for sample TF1 at 8 GHz at different temperatures (given in the figure) in zero dc magnetic field. The inset in Fig. 1a shows $T$-dependence of $R_s$ at low (1 kA/m) and high (7 kA/m) microwave field values at two different frequencies of 8 and 16 GHz. The data at 8 GHz are scaled to those at 16 GHz using $\omega^2$ scaling law. Scaling exponents are given in the figures.

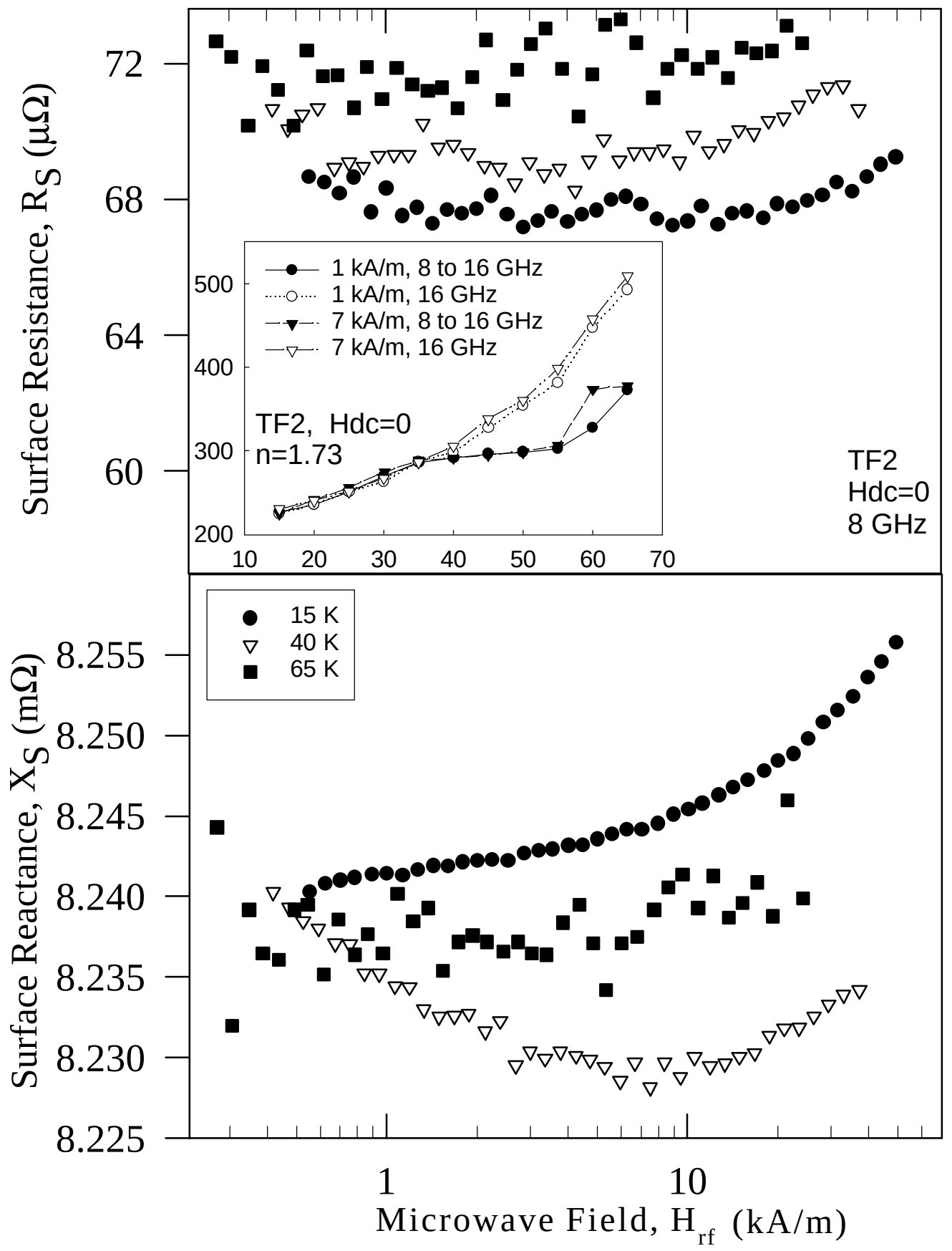

Fig. 2. Surface resistance, $R_s$ (a) and surface reactance, $X_s$ (b) as a function of microwave magnetic field, $H_{rf}$ for sample TF2 at 8 GHz at different temperatures (given in the figure) in zero dc magnetic field. The inset in Fig. 1a shows $T$-dependence of $R_s$ at low (1 kA/m) and high (7 kA/m) microwave field values at two different frequencies of 8 and 16 GHz. The data at 8 GHz are scaled to those at 16 GHz using $\omega^2$ scaling law. Scaling exponents are given in the figures.

model-independent way of determining absolute value of λ, provided that the film is homogeneous over its area.

### *B. Results and Discussion*

$H_{rf}$ -dependences of the surface resistance, $R_s$ and surface reactance, $X_s$ in zero dc magnetic field for films TF1 and TF2 are shown in Fig. 1 and Fig.2, respectively. Though both samples exhibit anomalies in either $R_s(H_{rf})$ or $X_s(H_{rf})$, the power dependences of $R_s(H_{rf})$ in general are rather different. For sample TF1, there is a well pronounced decrease in $R_s(H_{rf})$ at low temperatures ($T$ ~ 15 K), which gets "washed out" with increased $T$. $R_s(H_{rf})$ becomes fairly flat by ~ 40 K, and, finally, increases slightly at higher $T$. Here, $X_s(H_{rf})$ does not exhibit any anomalies over the whole temperature range studied (from 12 to 70 K).

For sample TF2, $R_s(H_{rf})$ exhibits a shallow minimum at low-$T$ (15 K), whereas $X_s(H_{rf})$ behaves in the usual way (i.e. it rises). However, with increased $T$, the anomalous behavior in $R_s(H_{rf})$ disappears, but quite a noticeable decrease in $X_s(H_{rf})$ develops (see Fig. 2, 40 K data). With further increase in $T$, anomalies in both $R_s(H_{rf})$ and $X_s(H_{rf})$ vanish (Fig.2 , 65 K).

The common feature for both the samples is that $R_s(H_{rf})$ and $X_s(H_{rf})$ *in zero dc field are always uncorrelated*. The major difference is that for TF1 only $R_s(H_{rf})$ shows anomalous behavior, whereas for TF2 it is mainly $X_s(H_{rf})$.

Another significant distinction between the samples is that for TF1 the anomalies in $R_s(H_{rf})$ are seen only at low $T$ (< 40 K, being most pronounced at the lowest $T$ of 15 K), whereas anomalies in $X_s(H_{rf})$ for TF2 develop only by 40-45 K, and are not seen at lower temperatures.

We have also estimated the frequency dependence of the surface resistance for the two samples by using the method of data collapse. Insets to Figs. 1a and 2a demonstrate the temperature dependences of $R_s$ for samples TF1 and TF2 in the low ($H_{rf}$ ~ 1 kA/m) and high ($H_{rf}$ ~ 7 kA/m) power regimes at two different frequencies, 8 and 16 GHz. Data obtained at 8 GHz are scaled to 16 GHz data by using a power-law approximation, $R_s \sim w^n$. It appears that the frequency scaling ($n$=2.54 and 1.73 for films TF1 and TF2, respectively) is somewhat different from $\omega^2$, which is usually expected in the low-power regime. Similar cases of deviation from the $\omega^2$-dependence have been reported in the literature (see [13] and references therein). Such a deviation is usually attributed to enhanced concentration of normal carriers and increased quasiparticle scattering time [13]. For the films studied in this work a different frequency dependence might be due to different microstructure of the samples, since film TF1 was fabricated at much a lower deposition temperature of 640 °C, as compared to the standard (for this method) deposition temperature of 690 °C for film TF2. Also, as AFM studies reveal, film TF1 exhibits a much smoother surface (average roughness is ~ 2 to 2.5 nm) compared to film TF1 (with the surface roughness of 3.5 to 5 nm).

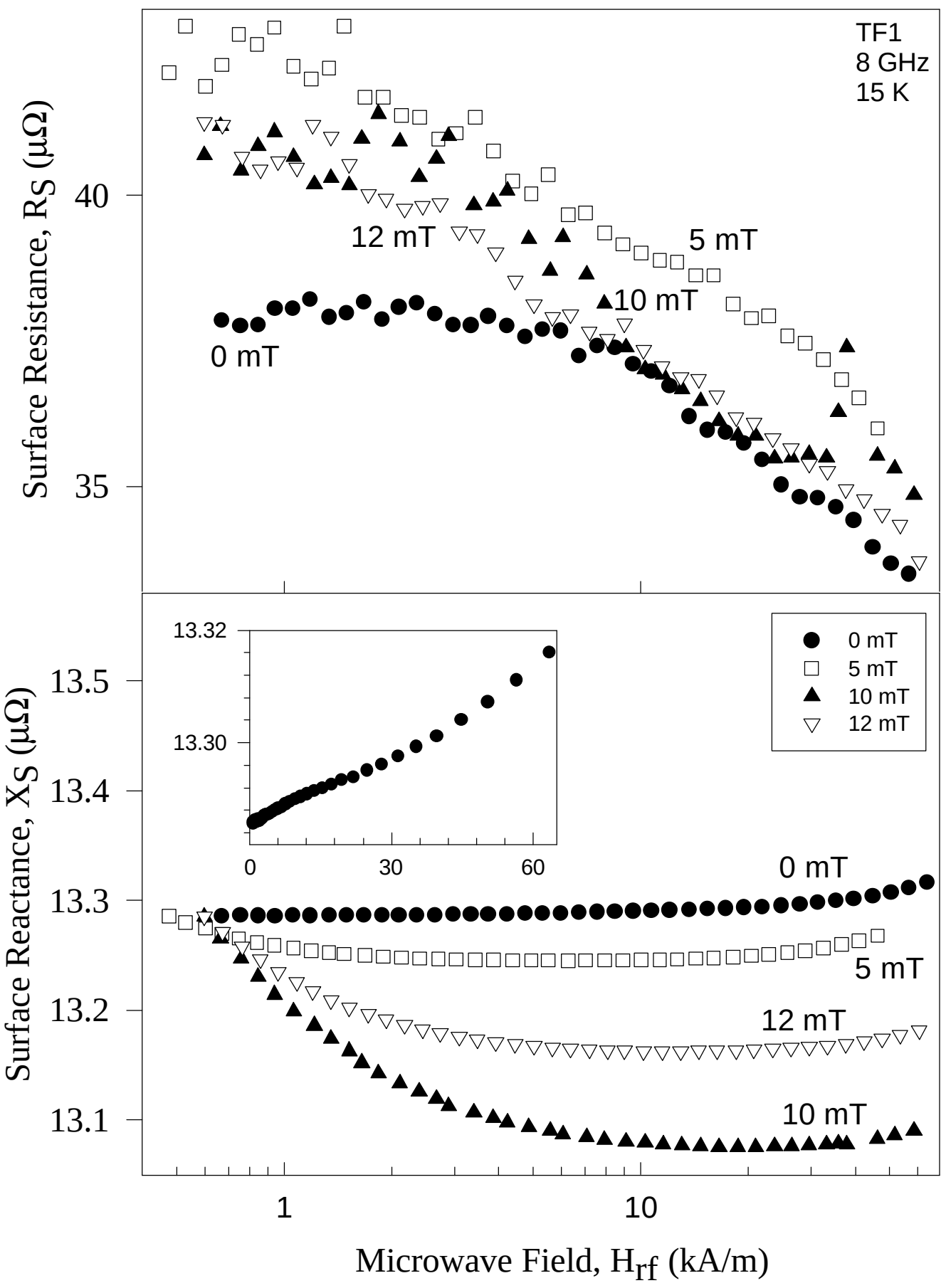

Fig. 3. Surface resistance, $R_s$ (a) and surface reactance, $X_s$ (b) as a function of microwave magnetic field, $H_{rf}$ for sample TF1 at 8 GHz at 15 K in finite applied dc magnetic field (given in the figures).

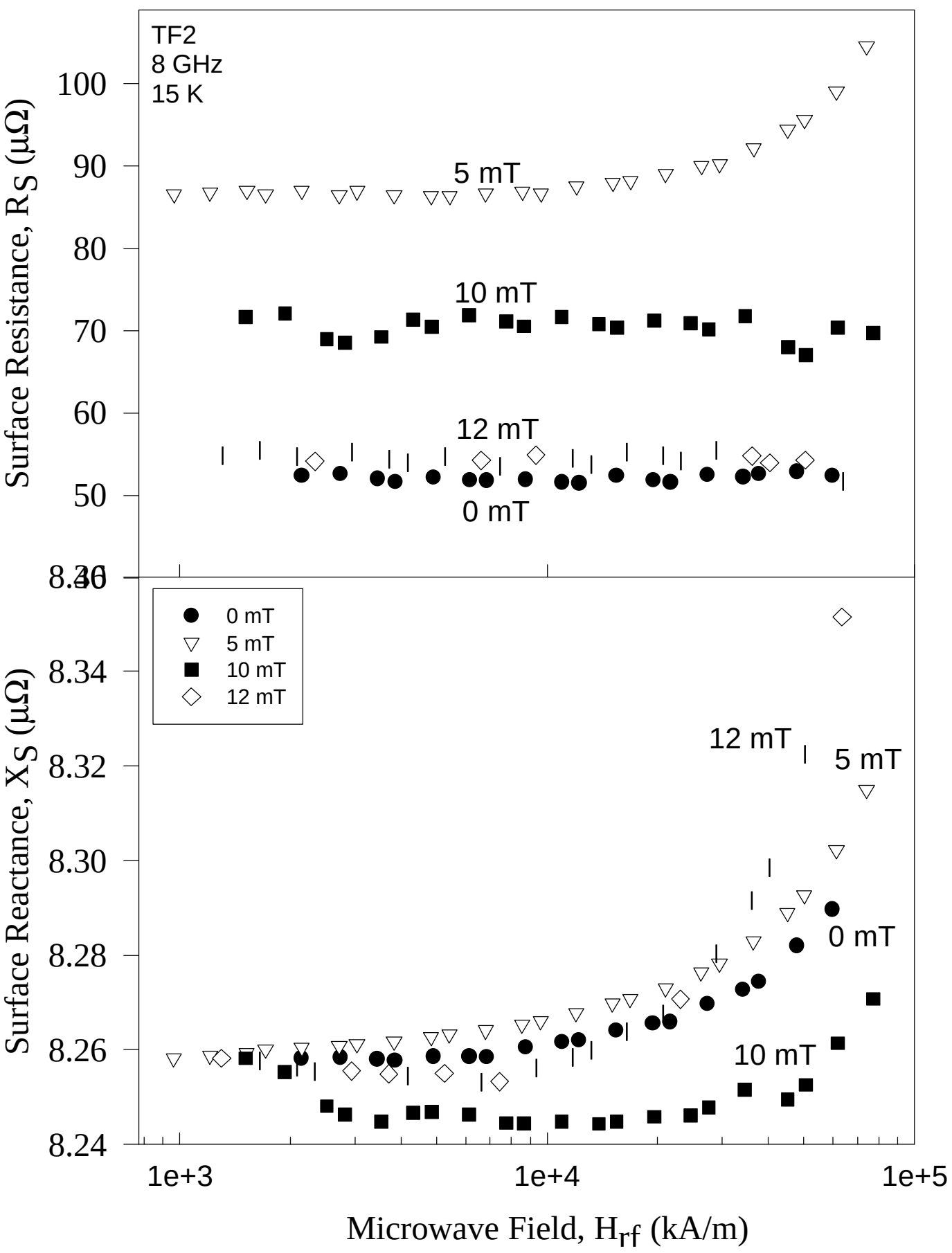

Fig. 4. Surface resistance, $R_s$ (a) and surface reactance, $X_s$ (b) as a function of microwave magnetic field, $H_{rf}$ for sample TF2 at 8 GHz at 15 K in finite applied dc magnetic field (given in the figures).

An important feature to be noticed here is that the same scaling exponent, $n$ holds in both the low and high microwave power regimes for both the samples. This appears to be at odds with high-field vortex dynamics observed by Belk et al. [14] in HTS films, which was attributed to the mechanism of vortex segments hopping between potential wells with a wide distribution of activation energies. This observation seems to suggest that low-field non-linear dynamics studied in our experiments is not affected by the vortex hopping mechanism.

The power dependences for samples TF1 and TF2 in various dc magnetic fields applied perpendicular to the sample surface (field-cooled regime) are shown in Fig.3 and Fig. 4. As one can conclude from the figures, the effect of relatively small dc magnetic fields on $Z_s(H_{rf})$ of HTS films can be dramatic. As far as sample TF1 is concerned, the dc field completely inverts $X_s(H_{rf})$, making it almost the "mirror image" of the zero field curve (see Fig.3b). Here, $R_s(H_{rf})$ is much less affected, with the main effect of $H_{dc}$ on $R_s(H_{rf})$ being an increase in low-power loss and, also, an enhanced steepness of the low-power portion of $R_s(H_{rf})$ (see Fig.3a). Here, the curves tend to merge at elevated values of $H_{rf}$.

With respect to sample TF2, it is seen that $H_{dc}$ can also induce anomalies in $X_s(H_{rf})$ (see Fig. 4a, 10 mT curve). Here, $R_s(H_{rf})$ is again affected to much less of an extent than $X_s(H_{rf})$, with the main result being an increase in the low-power $R_s$-value. Thus, a common feature for both the films is that the dependence of $Z_s$ on $H_{dc}$ at all $H_{rf}$ is non-monotonic, and $H_{dc}$ always enhances $R_s$, but at the same time can lead to a decrease in $X_s$ (see Figs. 3b and 4b).

## III. Conclusion

We have observed anomalous behavior of the non-linear microwave surface impedance, $Z_s$ of two high-quality Y-Ba-Cu-O thin films both in zero and finite applied dc magnetic fields at 8 and 16 GHz, using the coplanar resonator technique. For both of the samples the behavior of the surface resistance, $R_s$ and surface reactance, $X_s$ as a function of microwave field, $H_{rf}$ in zero dc field is uncorrelated. An anomalous decrease in $R_s(H_{rf})$ for film TF1 is observed only at low temperatures ($T \leq 40$ K), being more pronounced with reduced $T$. Conversely, for sample TF2 anomalies are observed only in $X_s(H_{rf})$ and are most noticeable at higher temperatures ($\geq 40$ K). Application of a weak (~ few mT) dc magnetic field, $H_{dc}$ can qualitatively change the power dependence of $X_s$ for both of the samples, but not that of $R_s$. In non-zero $H_{dc}$ correlated anomalous behaviors of $R_s(H_{rf})$ and $X_s(H_{rf})$ are observed.

The frequency dependence of $R_s$ for the two films remains the same in the low and high microwave power regimes, suggesting that the vortex segments hopping mechanism, first proposed by Belk et al. [14] for Y-Ba-Cu-O thin films at high dc magnetic fields ($\geq 0.5$ T) is irrelevant in our case.

Finally, while a small dc field usually enhances microwave losses, it is seen to decrease the absolute value of the penetration depth. In addition, the dependences of $R_s$ and $X_s$ on $H_{dc}$ are non-monotonic in both the low and high microwave power regimes.

The anomalous non-linear properties of HTS thin films reported in this paper have a number of important implications for microwave applications of these materials. First of all, since the anomalous effects are usually observed at rather low levels of the microwave (and/or dc) fields, they should be considered as detrimental to the power-handling capability of the thin films. In particular, manifestation of the anomalous behavior in $Z_s(H_{rf})$ should certainly give rise to appearance of the intermodulation distortion that is known to be extremely detrimental to the performance of various HTS passive microwave devices such as filters, for example. Appearance of the intermodulation products and harmonic generation will inevitably produce interference between adjacent channels in multi-channel communication systems. On the other hand, if those anomalous features prove to be adjustable by controlling the composition and deposition conditions for HTS films, we could hope to improve not only the power handling, but hopefully also the low-power $R_s$ values of the thin films and resultant microwave devices.

We are currently undertaking a research program that is aimed at establishing correlation between the non-linear surface impedance and intermodulation distortion in high-quality Y-Ba-Cu-O films, both in unpatterned and patterned states. The accompanying microstructural investigation of the samples with the help of various powerful techniques (AFM, magneto-optics, electron beam-induced voltage, modulated optical reflectance etc.) is hoped to shed more light on the origin of the anomalous non-linear effects and to assist in establishing the procedures of their successful control.

## Acknowledgment

A.V.V. thanks Dr. A. P. Kharel for assistance in performing part of the experiments during Dr. Kharel's Ph.D. project at the School of Electronic and Electrical Engineering, University of Birmingham.